\documentclass[onecolumn]{pasj00}

\begin{document}
\SetRunningHead{H. Takahashi et al.}
{ASCA Observations of the Central Regions of M31}
\Received{2001/04/26}
\Accepted{2001/08/21}

\title{ASCA Observations of the Central Regions of M31}
\author{
Hiromitsu \textsc{Takahashi}, Yuu \textsc{Okada},
Motohide \textsc{Kokubun}, and Kazuo \textsc{Makishima}}
\affil{Department of Physics, University of Tokyo,
7-3-1 Hongo, Bunkyo-ku, Tokyo 113-0033}
\email{hirotaka@amalthea.phys.s.u-tokyo.ac.jp}

\kword{galaxies: individual (M31) --- galaxies: spiral --- X-rays:
galaxies}

\maketitle

\begin{abstract}
Using ASCA, spatially integrated X-ray spectra of the central regions
of M31 were studied.
Data were accumulated over three different circular regions, with
the radii of $3'$, $6'$ and $12'$, all centered on the nucleus.
The spectra are relatively similar among the three regions.
In the energy range above 1.5 keV, the spectra are reproduced by a
combination of a disk black-body component and a black-body component,
implying that the emission mainly comes from an assembly of low-mass
X-ray binaries.
In energies below 1.5 keV, the spectra involves two additional softer
components, expressed with thin-thermal plasma emission models of
temperatures $\sim 0.9$ keV and $\sim 0.3$ keV.
Over the central $12'$ (2.4 kpc) region and in the 0.5--10 keV energy
band, the binary component has a luminosity of $2.6 \times 10^{39}$
erg s$^{-1}$, while the two softer components both exhibit luminosities
of $\sim 2 \times 10^{38}$ erg s$^{-1}$.
These results are compared with those from other missions, including
Chandra and XMM-Newton in particular.
Discussion is made on the nature of the two softer spectral components
besides the binary one.
\end{abstract}

\section{Introduction}

M31, the Andromeda Nebula, is the nearest ($\sim 700$ kpc) spiral
galaxy, forming the Local Group of galaxies together with our Milky Way
and others.
Its close distance enables us to perform very detailed observations in
the X-ray band.
Because the direction to M31 is clear of strong absorption along the
Galactic plane, it potentially allows more comprehensive X-ray
spectroscopic studies of X-ray sources over a wider energy range 
than is possible for those in our own Galaxy.

Consequently, many X-ray observations of M31 were so far performed.
The soft X-ray maps of M31 obtained by Einstein and ROSAT respectively 
revealed over 100 sources with the 0.2--4.0 keV luminosity 
exceeding $5 \times 10^{36}$ erg s$^{-1}$ (\cite{M31_Trinchieri}), 
and more than 300 sources with the 0.1--2.4 keV luminosity exceeding 
$3 \times 10^{35}$ erg s$^{-1}$ (\cite{M31_Supper}).
In the hard 2--20 keV energy band, the integrated spectrum of the
galaxy obtained with Ginga has been understood as emission
from a collection of X-ray binaries, with the integrated luminosity
reaching $\sim 5 \times 10^{39}$ erg s$^{-1}$ (Makishima et al.\ 1989b;
hereafter \authorcite{M31_Makishima}).

The Einstein and ROSAT observations also revealed unresolved and
apparently extended X-ray emission from the bulge of M31.
This is thought to be emission from a diffuse hot plasma, or from a
collection of faint discrete sources below the detection limit
(\cite{M31_Trinchieri,M31_Primini,M31_Supper,M31_West,M31_Irwin}b;
\cite{M31_diffuse_Borozdin}).
Using the BeppoSAX spectra, \citet{M31_diffuse_Trinchieri}
reconfirmed that the hard-band ($\gtrsim 2$ keV) emission from M31
can be described by a collection of binary sources, 
while an additional spectral component is required in the soft band.
However, due to the limited energy resolution in the soft band, 
the properties of the additional soft emission remains poorly
understood.
A further confusion as to the nature of the soft component is caused by
uncertainties in the way of modeling the X-ray binary spectra below
$\sim 1.5$ keV (e.g., \cite{M31_Irwin}b; \cite{M31_diffuse_Trinchieri}),
which persist even in the latest studies with Chandra and XMM-Newton
(\cite{M31_diffuse_Primini,M31_diffuse_Garcia}b;
\cite{M31_diffuse_Shirey}).

To study this unresolved emission, we analyzed the integrated
0.6--10 keV spectra of the central (up to 2.4 kpc) regions of M31 taken
with the GIS (Gas Imaging Spectrometer;
\cite{GIS_Ohashi,GIS_Makishima}) and the SIS (Solid-state Imaging
Spectrometer; \cite{SIS_Burke,SIS_Gendreau,SIS_Yamashita}), 
both onboard ASCA (\cite{ASCA_Tanaka}).
By employing a physical modeling of the X-ray binary spectra
established through observations of Galactic X-ray sources
(\cite{LMXB_Mitsuda,LMXB_Makishima}a; \cite{LMXB_Asai}),
we have discovered that the unresolved soft emission involves 
two plasma temperatures, $\sim 0.9$ keV and $\sim 0.3$ keV.
Therefore, the unresolved emission is likely to consist of two distinct
components.

\section{Observation and Data Reduction}

\begin{figure}
\begin{center}
\FigureFile(80mm,80mm){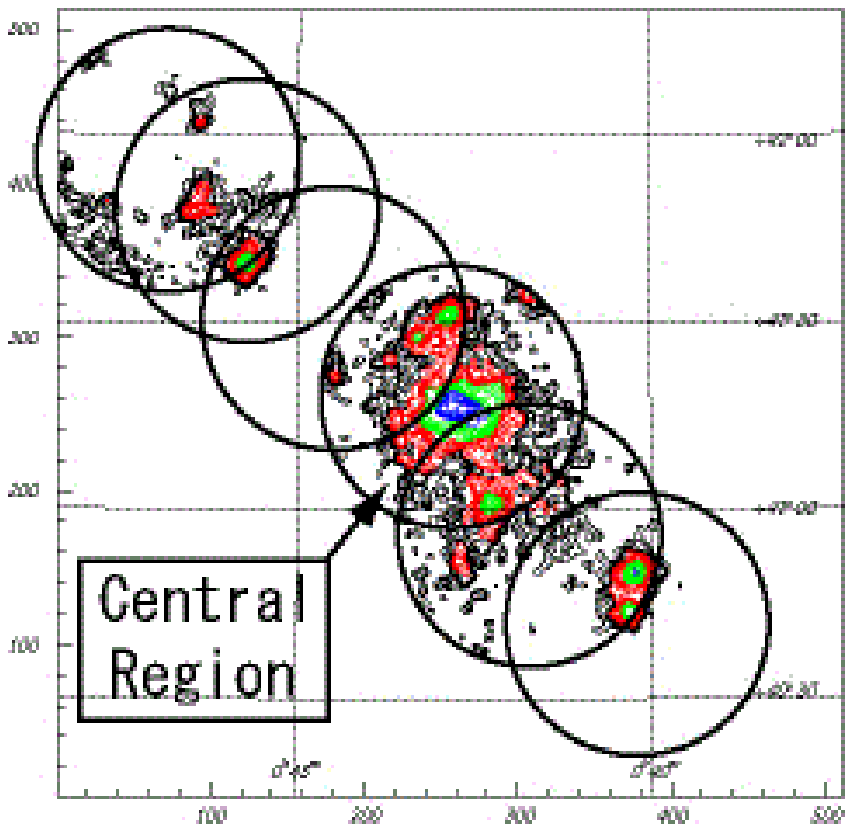}
\end{center}
\caption{The GIS2 images of six pointings onto M31, in the energy range
of 0.7--10 keV.
Each of the six circles represents the field of view of the GIS one
pointing.
The background is included, and the contours are logarithmically spaced.
Sky coordinates are J2000.
The present paper is concerned only with the central pointing.}
\label{fig:M31_image}
\end{figure}

The ASCA observations of M31 were performed from 28 to 31 July 1993 
in the performance verification phase for $\sim 120$ ks in total.
The observations were divided into six partially-overlapping pointings
covering a large fraction of the optical galaxy disk, including the
central region in particular.
During all these observations, the GIS was operated in normal
pulse-height mode, and the SIS was in 4CCD mode.
In figure~\ref{fig:M31_image}, we show the mosaic GIS images from all
the six pointings.
Thus, the central region of M31 is too crowded to be resolved by ASCA.
In the present paper, we therefore analyze the data from this
central-pointing field as a collection of many discrete sources and
possibly diffuse emission, in an attempt to separate different X-ray
components spectrally rather than spatially.

For the central pointing data, we screened the events under the
condition of a telescope viewing direction of $> 5^{\circ}$ from the
dark Earth rim, and a magnetic cutoff rigidity $> 8$ GV and $> 6$ GV
for the GIS and the SIS, respectively.
For the SIS data, we further required the time after day-night
transition to be $> 100$ s, and the elevation angle from the sunlit
Earth rim to be $> 30^{\circ}$ in order to avoid the effect of light
leakage on the CCD chips.
We also removed particle events by the standard rise-time rejection for
the GIS, and similarly removed hot or flickering pixels for the SIS.
After applying these criteria, we obtained net exposures of $\sim 21$
ks with the GIS (GIS2+GIS3) and $\sim 14$ ks with the SIS (SIS0+SIS1).
The 0.7--10 keV count rates were $\sim 0.9$ c s$^{-1}$ 
and $\sim 1.1$ c s$^{-1}$, for the GIS and SIS, respectively.


\section{Data Analysis and Results}

\subsection{Accumulation of Energy Spectra}

After the data screening described in section 2, we constructed the GIS
and SIS energy spectra for the central region of M31.
The events were extracted from a circular region centered on the
nucleus, using a radius of $12'$ (2.4 kpc), the largest possible radius
without being affected by dominant point sources which should be
analyzed individually.
As to the SIS, the field of view ($22' \times 22'$ in 4CCD mode) is a
little smaller than the $12'$ radius, and we use only the region on the
detector.
We derived background data from blank-sky observations, 
for which we applied the same data-selection criteria as in section 2, 
and used the same photon-accumulation region.
After the background subtractions and appropriate gain corrections, 
we added the data from GIS2 and GIS3 into a single GIS spectrum, 
and those from SIS0 and SIS1 into a single SIS spectrum.
The obtained spectra are shown in figure~\ref{fig:1.5_10keV} 
without removing the instrumental responses.
The signal X-rays are thus detected over the 0.7--10 keV range with the
GIS, and over the 0.6--7.5 keV range with the SIS.

In order to roughly investigate the spatial distribution of X-ray
sources, we also accumulated two more pairs of spectra over smaller
circular regions centered on the nucleus, using radii of $3'$ (0.6 kpc)
and $6'$ (1.2 kpc).
The former radius is twice as large as the half-power radius of ASCA.
Below, we analyze the $12'$ spectra in subsections 3.2 through 3.4,
and apply the results to the $3'$ and $6'$ ones in subsection 3.5.


\subsection{Modeling of the 1.5--10 keV Spectra}

The X-ray emission from M31 is thought to be dominated by a collection
of low-mass X-ray binaries (LMXBs, accreting binaries involving neutron
stars with weak magnetic fields) at least in energy bands above
$\sim 1.5$ keV (\authorcite{M31_Makishima};
\cite{M31_diffuse_Trinchieri, M31_diffuse_Shirey}).
Indeed, when the absorption was fixed to the Galactic value along the
line of sight toward M31, $N_{\rm H} = 6.7 \times10^{20}$ cm$^{-2}$
(from Einline and W3nH), the 2--20 keV spectrum of the whole M31
obtained with Ginga has been described successfully
(\authorcite{M31_Makishima}) by the physical model developed for
Galactic high-luminosity LMXBs; it consists of a disk black-body (DBB)
component and a black-body (BB) component, which represent emission
from the optically-thick accretion disk and the central neutron star,
respectively (\cite{LMXB_Mitsuda,LMXB_Makishima}a; \cite{LMXB_Asai}).
We hereafter refer to this model as the LMXB model (see Appendix for
detail). 
In contrast, the Ginga spectrum was not reproduced by thermal
Bremsstrahlung or power-law models, unless a significant amount of
excess absorption was incorporated. 
Since the absorption within M31 itself is rather small, the excess
absorption is considered artificial, making these two alternative
spectral models unrealistic.

We accordingly fitted the 1.5--10 keV GIS and SIS spectra of the
central $12'$ radius simultaneously with the LMXB model.
Here and hereafter, we fix the value of the absorption column density
at the Galactic value.
As summarized in table~\ref{tab:LMXB}, the LMXB model has
simultaneously reproduced the GIS/SIS spectra very well 
in the energy range above 1.5 keV, and the obtained parameters are
consistent with those obtained previously with Ginga
(\authorcite{M31_Makishima}).
Furthermore, the inner-disk temperature of $kT_{\rm in} \sim 1.0$ keV,
the BB temperature of $kT_{\rm BB} \sim 2.0$ keV, and the cross-over of
the two components at 3.3 keV (figure~\ref{fig:1.5_10keV}), 
are all typical of luminous Galactic and Magellanic LMXBs
(\cite{LMXB_Mitsuda}).
It is therefore reconfirmed that a collection of LMXBs dominates the
X-ray emission from M31 above $\sim 1.5$ keV.
In subsection 3.6, we examine what happens if our LMXB model is
replaced with more conventional ones, such as a single thermal
Bremsstrahlung or a single power-law model, even though these
alternative modelings have already been ruled out by Ginga
(\authorcite{M31_Makishima}).


\subsection{Inclusion of the 0.8--1.5 keV Energy Band}

\begin{figure}
\begin{center}
\FigureFile(80mm,80mm){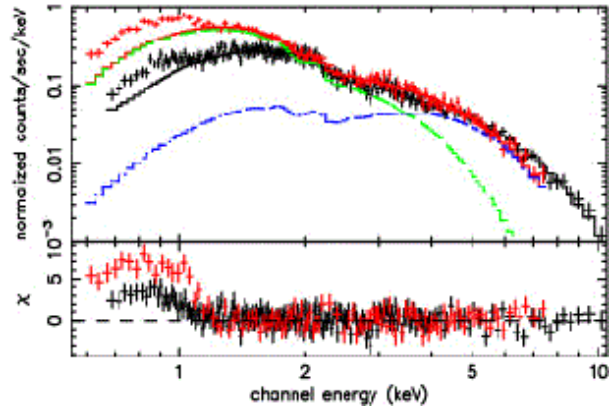}
\end{center}
\caption{Background-subtracted GIS (black) and SIS (red) spectra of
the central $12'$ region of M31, presented without removing the
instrumental response.
They are fitted simultaneously in the energy range above 1.5 keV,
with the canonical LMXB model.
The green and blue lines represent the DBB and BB model components
contributing to the SIS spectrum, respectively.
The best fit model is extrapolated to below 1.5 keV, to highlight the
soft excess.}
\label{fig:1.5_10keV}
\end{figure}

When the best-fit LMXB model determined in the 1.5--10 keV band is
extrapolated toward lower energies, the model prediction falls
significantly short of the actual data (figure~\ref{fig:1.5_10keV});
the fit actually becomes unacceptable by including the 0.8--1.5 keV
band, even re-adjusting the model parameters except $N_{\rm H}$
(table~\ref{tab:LMXB}).
This reconfirms the previous reports on the soft X-ray excess
(\cite{M31_Irwin}b; \cite{M31_diffuse_Trinchieri,M31_diffuse_Borozdin,
M31_diffuse_Shirey}).
The soft excess is consistently seen between the two ASCA instruments, 
although it is more significant in the SIS data because of its improved
low-energy efficiency.

\begin{figure}
\begin{center}
\FigureFile(80mm,80mm){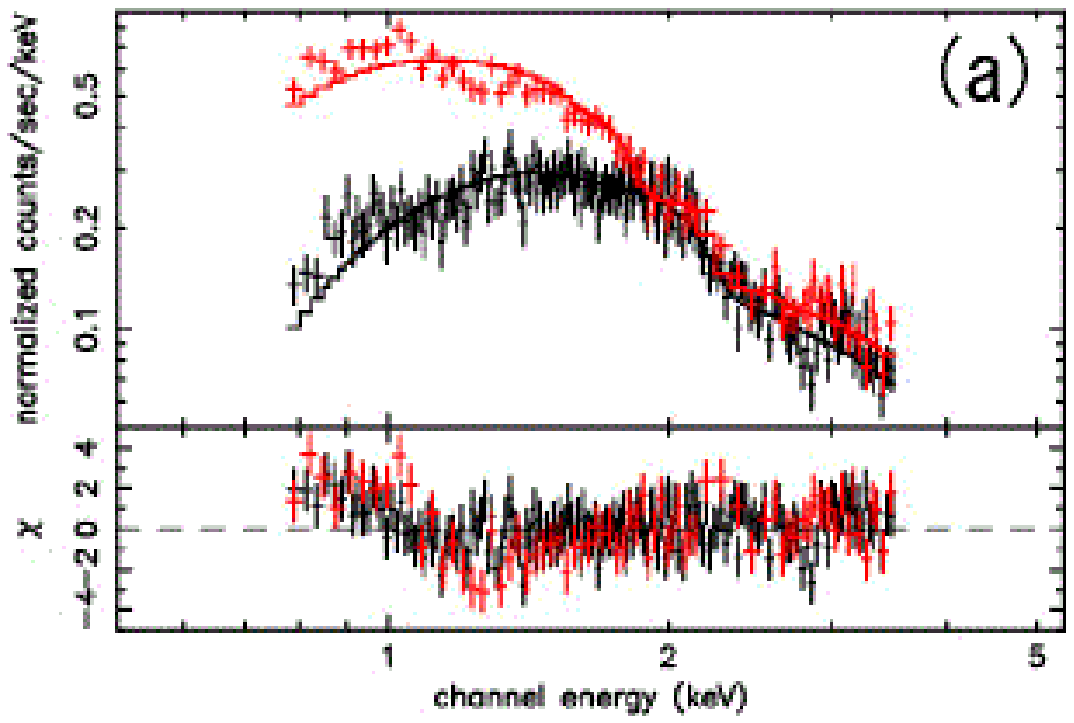}
\FigureFile(80mm,80mm){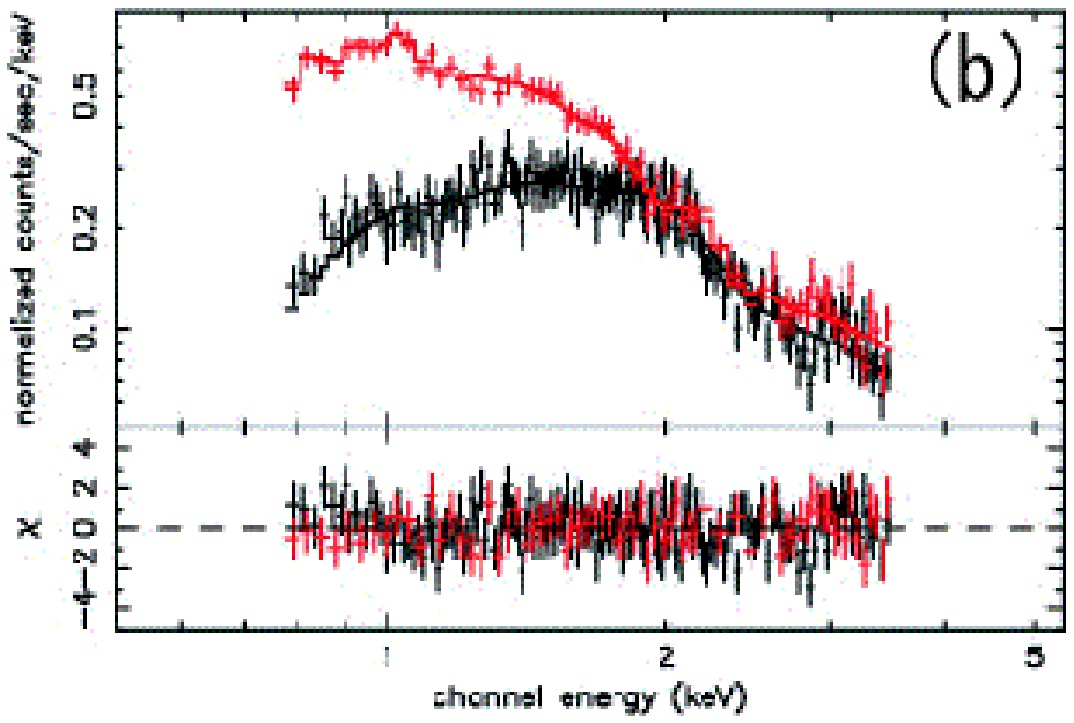}
\FigureFile(80mm,80mm){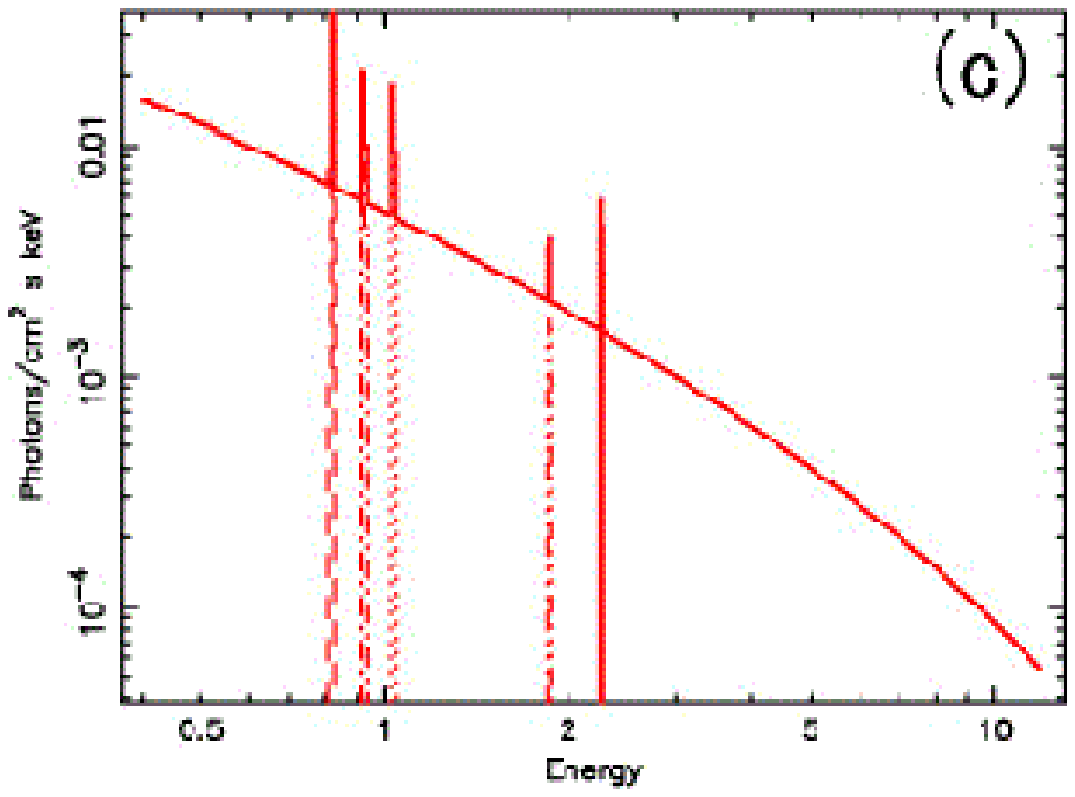}
\end{center}
\caption{Spectra of the central $12'$ region of M31,
presented over a restricted 0.8--3.5 keV energy band.
The data are the same between the two panels.
(a) Fitted jointly with a single bremsstrahlung model.
(b) Five narrow Gaussians are added,
to represent ionized Ne-K, Fe-L, Si-K, and S-K lines.
The line parameters are given in text.
(c) The incident model corresponding to the fit in panel b.}
\label{fig:line}
\end{figure}

We suspect the soft excess, prominent below energies of $\sim 1$ keV in
figure~\ref{fig:1.5_10keV}, to be contributed significantly by
low-energy atomic lines emergent from thin-thermal plasmas.
We accordingly examined the 0.8--3.5 keV portion of the GIS/SIS spectra
more closely for the existence of atomic emission lines from major
elements.
We approximated the overall GIS/SIS continua (including the soft excess)
conventionally by a single Bremsstrahlung model, to find a rather poor
fit with $\chi^2$/d.o.f.\ = 351/227 as shown in figure \ref{fig:line}a. 
By including 5 Gaussians, in contrast, the fit has been significantly
improved ($\chi^2$/d.o.f.\ = 211/217; figure \ref{fig:line}b).
The five Gaussians are centered at $0.82^{+0.02}_{-0.01}$ keV, 
$0.93^{+0.01}_{-0.02}$ keV, $1.03^{+0.01}_{-0.02}$ keV, 
$1.85^{+0.06}_{-0.05}$ keV, and $2.26 \pm 0.04 $ keV, with equivalent
widths of $49^{+10}_{-12}$ eV, $36^{+10}_{-13}$ eV, $38^{+11}_{-8}$ eV,
$15 \pm 11$ eV, and $44^{+11}_{-25}$ eV, respectively.
Thus, the lines are statistically significant.
Based on the central energies, the former three lines may be assigned
to either ionized Ne-K or ionized Fe-L lines, while the latter two to
ionized Si-K and S-K lines, respectively.
We can hence conclude that the spectra exhibit statistically
significant evidence of low-energy emission lines from ionized
abundant heavy elements.

Now that the thin-thermal nature of the soft excess has been confirmed,
we are justified to express the soft excess by a thin-thermal plasma
emission model. 
Specifically, we employ the Raymond-Smith emission model
(Raymond \& Smith 1977; hereafter RS model), and tentatively fix its
metal abundance  at 1.0 solar, as has been found from the Galactic
ridge emission (\cite{ridge_Kaneda,ridge_Valinia}) and an apparently
diffuse X-ray emission filling the Galactic bulge
(\cite{bulge_Kokubun}).
We hence fitted the GIS/SIS spectra over the 0.8--10 keV energy band 
by the sum of the LMXB and RS components.
This two-component model, denoted LMXB+RS model, has successfully
explained the 0.8--10 keV GIS/SIS spectra, with the plasma temperature
being $\sim 0.8$ keV (table \ref{tab:LMXB}).
In other words, the soft excess seen in figure \ref{fig:1.5_10keV}
below 1.5 keV has been explained successfully as emission from a
0.8 keV thin-thermal plasma.


\subsection{Inclusion of the 0.6--0.8 keV Energy Band}

\begin{figure}
\begin{center}
\FigureFile(80mm,80mm){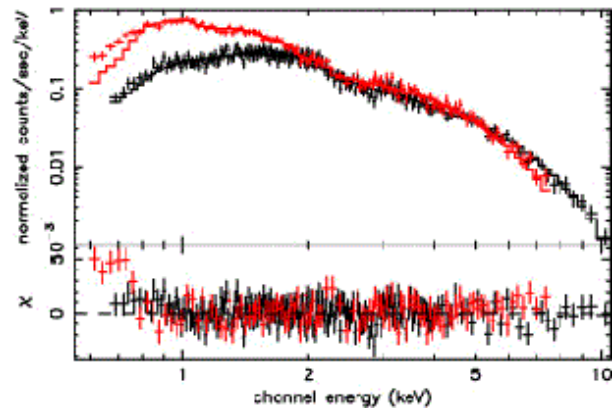}
\end{center}
\caption{The same as figure~\ref{fig:1.5_10keV}, but the simultaneous
model fits are performed over the 0.8--10 keV energy range using the
LMXB+RS model.
The best fit model is extrapolated to below 0.8 keV, to reveal yet
another soft excess component.}
\label{fig:0.8_10keV}
\end{figure}

As in subsection 3.3, we extended the LMXB+RS model, determined over
the 0.8--10 keV energy band, further into the 0.6--10 keV energy band.
Then, as shown in figure~\ref{fig:0.8_10keV}, the model has again left
a large excess below $\sim 0.8$ keV.
As a result, the LMXB+RS fit that was acceptable in the 0.8--10 keV
range becomes unacceptable in the 0.6--10 keV range, even re-adjusting
the model parameters except $N_{\rm H}$ and abundances
(table~\ref{tab:LMXB}).
This indicates the presence of a third, and the softest, emission
component that dominates the spectra below $\sim 0.8$ keV.

\begin{figure}
\begin{center}
\FigureFile(80mm,80mm){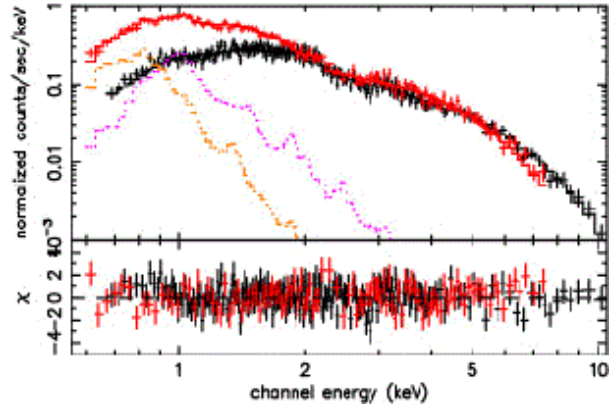}
\end{center}
\caption{The same as figure~\ref{fig:1.5_10keV},
but the model fits are performed over the full 0.6--10 keV energy range
using the LMXB+2RS model.
The pink and orange lines represent the 0.9 keV and 0.3 keV
RS model components contributing to the SIS spectrum, respectively.}
\label{fig:0.6_10keV}
\end{figure}

The softest component may be contributed by O-K lines appearing in the
0.6--0.7 keV range.
We therefore added one more RS component to jointly fit the total-band
(0.6--10 keV) GIS/SIS spectra.
This model, denoted LMXB+2RS model, has indeed decreased $\chi^2$ down
to an acceptable level, 
as shown in figure~\ref{fig:0.6_10keV} and table~\ref{tab:LMXB}.
The obtained two plasma temperatures are $\sim 0.9$ keV and $\sim 0.3$
keV, and their luminosities are given in table~\ref{tab:luminosity};
here and hereafter, we quote luminosities in the 0.5--10 keV energy
band.
Even if the metal abundances of the two RS components are allowed to
vary freely, the fit result does not change significantly, and neither
of the two RS components vanishes.
If, for simplicity, we constrain the two RS components to have common
abundances with solar ratios, the 90\% confidence range of the metal
abundance becomes $\geq 0.09$ solar.
When the MEKAL model (\cite{MEKALI_Mewe}, 1986;
\cite{MEKAL_Kaastra,MEKAL_Liedahl}) is used instead of the RS model,
the spectra also require two thin-thermal components besides the LMXB
component, and the obtained parameters are consistent, at the 90\%
confidence level, with those derived by using the RS model.


\subsection{Spectra of the Central $3'$ and $6'$ Regions}

\begin{figure}
\begin{center}
\FigureFile(80mm,80mm){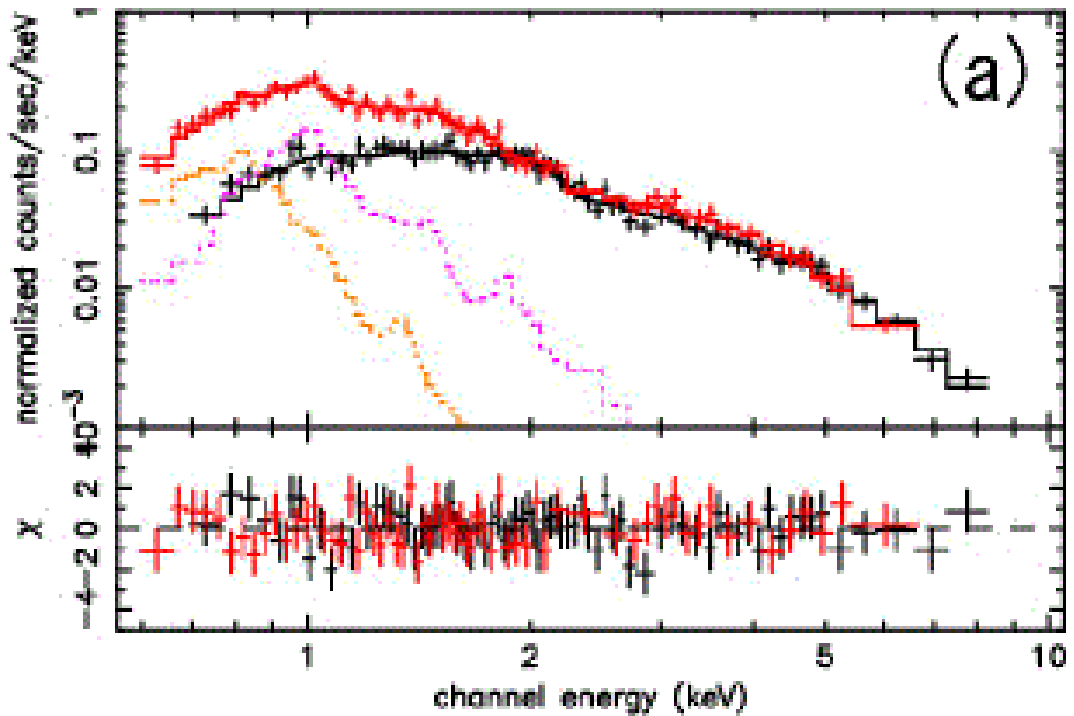}
\FigureFile(80mm,80mm){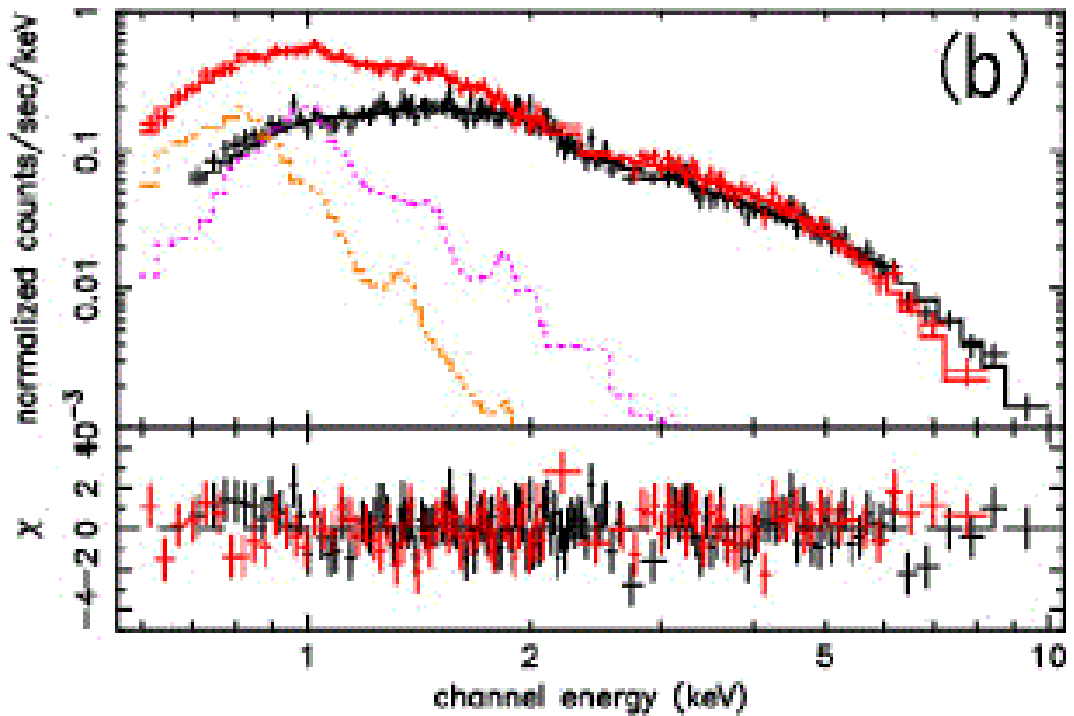}
\end{center}
\caption{Spectra of the central $3'$ (panel a) and $6'$ (panel b)
regions fitted with the LMXB+2RS model over the full 0.6--10 keV energy
band, presented in the same way as figure~\ref{fig:0.6_10keV}.}
\label{fig:3'+6'}
\end{figure}

In order to examine the reality and the nature of the three
spectroscopic components found above, it would help to study their
spatial distributions utilizing the imaging capability of ASCA.
Accordingly, we have accumulated the GIS and SIS events over two
concentric circular regions of smaller radii of $3'$ (0.6 kpc) and $6'$
(1.2 kpc), centered on the nucleus, as performed in subsection 3.1 for
the $12'$ region.
We again subtracted the blank-sky background spectra.
As presented in figure~\ref{fig:3'+6'}, the GIS/SIS spectra accumulated
over these regions are similar in shape to those derived from the $12'$
region (figure~\ref{fig:0.6_10keV}), and have actually been reproduced
successfully by the LMXB+2RS model.
The derived parameters, summarized in table~\ref{tab:3'+6'}, are close
to those found in the $12'$ accumulation region (table~\ref{tab:LMXB}).

In table~\ref{tab:luminosity}, we show the spatially-dependent
luminosities of the three spectral components constituting the
LMXB+2RS model.
Thus, the three components are statistically significant in every pair
of spectra.
All the three components are clearly extended, but are more concentrated
toward the nucleus than is expected for a uniform distribution.
While the luminosities of the LMXB component and the softest (0.3 keV
RS) component are consistent with having the same spatial distribution, 
there is a marginal evidence that the hotter (0.9 keV RS) component is
less extended around the nucleus.
All these results reinforce the reality of the three spectral
components.


\subsection{Alternative Modelings of the X-ray Binary Component}

Although we have thus successfully decomposed the spatially-integrated
M31 spectra, the LMXB model employed to represent the X-ray binary
contribution had been developed for the most luminous LMXBs with
luminosity exceeding $\sim 2 \times 10^{37}$ erg s$^{-1}$
(\cite{LMXB_Mitsuda,LMXB_Makishima}a; \cite{LMXB_Asai}), 
residing in our Galaxy and the Large Magellanic Cloud (specifically,
LMC~X-2).
Less luminous LMXBs tend to show more power-law like spectra due to
Comptonization (e.g., \cite{power_Mitsuda}).
Furthermore, there is a claim, though without physical grounds, 
that LMXBs might exhibit a prominent soft excess component of which the
strength is correlated with environmental metallicity
(\cite{LMXBI_Irwin,LMXBII_Irwin}ab).
With these in mind, we re-fitted the 0.6--10 keV GIS/SIS spectra for
the $12'$ region, but this time by replacing the LMXB model with a
single thermal Bremsstrahlung (Bremss for short) model or a single
power-law (PL for short) model.
In logarithmic plots, the Bremss model is more convex than the PL model,
but less convex than the LMXB model (see Appendix for detail).
We again fix the absorption to the Galactic value.

\begin{figure}
\begin{center}
\FigureFile(80mm,80mm){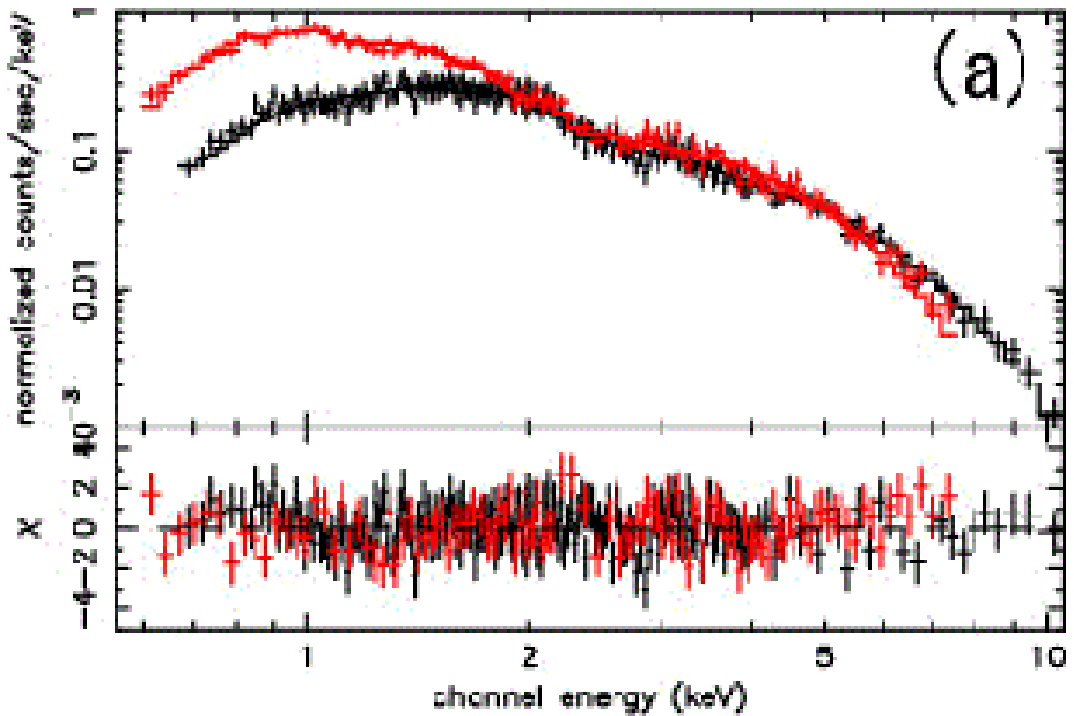}
\FigureFile(80mm,80mm){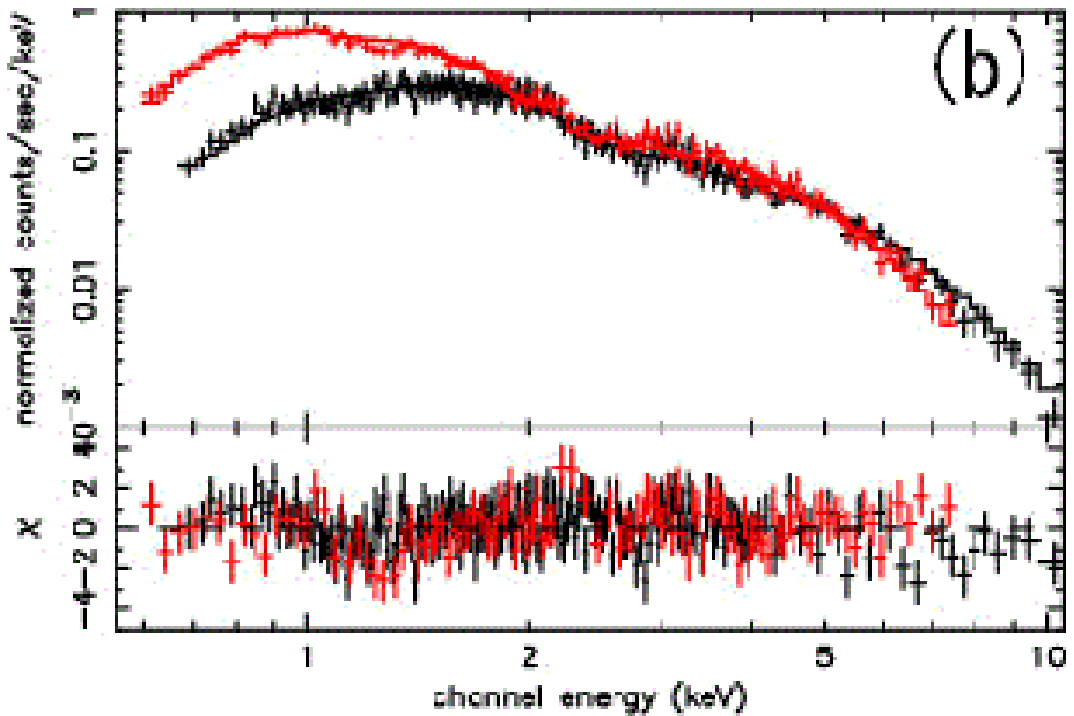}
\end{center}
\caption{The same as figure~\ref{fig:0.6_10keV},
but fitted with (a) the Bremss+2RS model, and (b) the PL+2RS model.}
\label{fig:bremss+power}
\end{figure}

As summarized in table~\ref{tab:bremss+power}, the single Bremss model
could not reproduce the 0.6--10 keV GIS/SIS spectra due to the soft
excess.
Adding an RS component (again with the abundance fixed at 1.0 solar)
improved the fit significantly, 
and further adding a second RS component has made the fit acceptable; 
the obtained Bremss+2RS fit is shown in figure~\ref{fig:bremss+power}a.
These results, together with the obtained RS parameters, are basically
similar to those we obtained using the LMXB model in
table~\ref{tab:LMXB}.
This is understandable, because the Bremss model of the temperature
$kT_{\rm B} \sim 10$ keV is similar in shape to the LMXB model over
the ASCA band, except that the former predicts a slightly higher flux
below $\sim 1.5$ keV.
However, the LMXB+2RS model reproduces the overall data better than the
Bremss+2RS model (with $\chi^2$/d.o.f.\ = 300/309 vs.\ 307/311), and the
difference is significant at a $> 97$\% confidence level according to
an $F$-test.
These results are consistent with the conclusion of
\authorcite{M31_Makishima} that the 2--20 keV spectrum of M31 is
represented significantly better by the LMXB model than the Bremss
model.

As to the PL modeling, the fit remained unacceptable even adding up to
two RS components (table~\ref{tab:bremss+power}).
The unsuccessful PL+2RS fit is shown in figure~\ref{fig:bremss+power}b,
where both instruments consistently reveal a negative residual feature
near 1.2 keV.
This is because the PL model overpredicts the continuum at this energy.
In short, the PL model is inappropriate 
as a representation of the X-ray binary spectrum in M31, as already
revealed in \authorcite{M31_Makishima}.


\section{Discussion}

Using the ASCA data, we have confirmed that the spatially
integrated 0.6--10 keV X-ray emission from the central regions of M31
comprises three spectral components; the integrated X-ray binary
component, and the 0.9 keV (hotter) and the 0.3 keV (cooler) RS
components.
All the three components are spatially extended, in agreement with the
repeated detections of unresolved, possibly diffuse, soft X-ray
emission (\cite{M31_Trinchieri,M31_Primini,M31_Supper,M31_West,
M31_diffuse_Trinchieri,M31_diffuse_Borozdin,M31_diffuse_Primini,
M31_diffuse_Garcia}b; \cite{M31_diffuse_Shirey}).
Below, we compare our results with those from other missions,
and discuss the nature of the three components.

\subsection{The Binary Component}

The integrated binary component dominates the spectrum above $\sim 1.5$
keV, as already known previously (e.g., \authorcite{M31_Makishima};
\cite{M31_diffuse_Trinchieri}).
We have successfully modeled this emission using our physical LMXB
model, consisting of a DBB component and a BB component.
This result agrees with the main achievement of
\authorcite{M31_Makishima}, 
that the 2--20 keV Ginga spectrum of the whole M31 is described
successfully using the same model.
Our parameters for the LMXB component are consistent with those derived
in \authorcite{M31_Makishima}.
Furthermore, when integrated up to $12'$, the 0.5--10 keV luminosity of
this component measured with ASCA ($2.6 \times 10^{39}$ erg s$^{-1}$;
table~\ref{tab:luminosity}) compares reasonably with the 2--20 keV
luminosity of the whole M31 derived with Ginga,
$5 \times 10^{39}$ erg s$^{-1}$ (\authorcite{M31_Makishima}).

We are of course aware that our LMXB model, which has primarily been
developed for very luminous ones, may not necessarily be appropriate
for fainter sources which must contribute to the spatially integrated
ASCA spectra.
The spectra of such fainter LMXBs, with luminosities below
$\sim 1 \times 10^{37}$ erg s$^{-1}$, may be described better by a PL
model with a mild high-energy turn over (\cite{power_Mitsuda}).
Nevertheless, the luminosity function of discrete X-ray sources 
in M31 obtained with ROSAT (\cite{M31_West}) indicates that the
integrated X-ray luminosity of M31 is predominantly contributed
by sources more luminous than $\sim 1 \times 10^{37}$ erg s$^{-1}$, 
with the fainter ones contributing only 10\% or less.
Furthermore, M31 does not host apparent examples of enigmatic
``ultra-luminous compact X-ray sources'' (ULXs; \cite{ULX_Makishima})
that shine extremely beyond the Eddington limit for a neutron star.
Consequently, the largest contribution to the total X-ray emission is
thought to come from LMXBs, justifying the use of our LMXB spectral
model.

In contrast to our approach, other investigators mostly model the
binary component in more conventional ways, in terms of either the
Bremss or the PL model (\cite{M31_diffuse_Trinchieri,M31_Irwin}b;
\cite{M31_diffuse_Borozdin,M31_diffuse_Shirey}).
The Bremss modeling may be consistent with our ASCA data, but evidently
this has no physical grounds.
Furthermore, the Bremss model has been rejected by the 2--20 keV Ginga
spectrum of the whole M31 (\authorcite{M31_Makishima}),
presumably because this energy band is more suited than the ASCA band 
to characterize the overall spectral shape of luminous point X-ray
sources.
The PL modeling has been revealed to be inconsistent with the
high-quality ASCA data.
We thus conclude that our LMXB model gives the most appropriate
account of the integrated binary emission from M31.

To confirm the above idea, we simulated the XMM-Newton EPIC spectrum
for bright point sources in M31 (blue one in fig.~7 of
\cite{M31_diffuse_Shirey}), employing the best-fit PL model of photon
index 1.82 and absorption $N_{\rm H} = 6.7 \times 10^{20}$ cm$^{-2}$
as reported by \citet{M31_diffuse_Shirey}, and using a publicly
available EPIC response (mos1\_medium\_all\_qe17\_rmf3\_tel5\_15.rsp).
When fitted with our LMXB model over the 0.6--7.0 keV range, the
simulated spectrum exhibits a weak soft excess below $\sim 0.8$ keV, 
yielding $\chi^{2}$/d.o.f.\ = 446/226.
This meets our expectation, since the spectrum must inevitably contain
a small amount of background/foreground diffuse emission, even though
it was accumulated over small radii ($10''-30''$;
\cite{M31_diffuse_Shirey}) around bright point sources.
The soft excess thus revealed is modeled reasonably well
($\chi^{2}$/d.o.f.\ = 288/262) by two RS components of one solar
abundances, with temperatures of $0.91^{+0.18}_{-0.05}$ keV and
$0.23^{+0.02}_{-0.03}$ keV.
When normalized to the binary component, the inferred luminosities of
the two RS components are $\sim 30$\% of those found from the ASCA
spectra.
This is reasonable, because the data integration area used to generate
the EPIC point-source spectrum is estimated to have a filling factor of
10--40\%, as judging from \citet{M31_diffuse_Shirey}.

\subsection{The Hotter RS Component}

The second component, the $\sim 0.9$ keV thin-thermal plasma emission, 
is a new component which has not been detected so far by other missions.
For example, the Chandra and XMM-Newton observations clearly revealed
the presence of apparently diffuse X-ray emission in the central region
of M31 (\cite{M31_diffuse_Primini,M31_diffuse_Garcia}b;
\cite{M31_diffuse_Shirey}).
\citet{M31_diffuse_Shirey} decomposed the spectrum of this unresolved
emission obtained by XMM-Newton into a soft thin-thermal plasma
component of a temperature $\sim 0.35$ keV, and a hard residual emission
from fainter discrete sources.
However, they did not find the 0.9 keV thin-thermal plasma which we have
discovered.

This discrepancy, we believe, is again due to the difference in the way
of modeling the residual binary component contributing to the unresolved
emission.
\citet{M31_diffuse_Shirey} modeled the residual component in terms of
the PL model, which as well as the Bremss is steeper in the soft X-ray
range than our LMXB model (Appendix).
As a result, the 0.9 keV component in the diffuse emission was
presumably taken up by the model describing the binary contribution.
Actually, when we replace the LMXB model by the Bremss model in the
ASCA spectral fit, the normalization (and hence luminosity) of the 0.9
keV component decreases to $\sim 70$\% of its original value, although
it does not vanish.
Furthermore, when the binary contribution to the ASCA spectra is
represented by a PL model, the normalization of the 0.9 keV component
reduces to $\sim 40$\% of that found with the LMXB modeling (the fit
becoming unacceptable; subsection~3.6, figure~7b, table~4).
By Chandra with its excellent imaging capability, the same region was
resolved into more point sources than those by XMM-Newton, and
the contribution of them to the unresolved emission can be further
reduced.
However, there have not yet been any detailed spectral analyses but for
a hardness ratio one which confirmed the value is consistent with that
expected from a $\sim 0.3$ keV RS component
(\cite{M31_diffuse_Primini,M31_diffuse_Garcia}b).

We can independently strengthen the reality of the 0.9 keV component, 
by considering atomic emission lines.
As described in subsection 3.3, the ASCA spectra bears a statistically
significant ionized S-K line; so does the diffuse-component spectrum
obtained with the XMM-Newton EPIC (red one in fig.~7 of
\cite{M31_diffuse_Shirey}).
Since the S-K lines cannot be emitted significantly by plasmas of
temperatures as low as 0.3--0.4 keV, the data require a considerably
higher plasma temperature.
Furthermore, a close look at the same XMM-Newton EPIC spectrum for the
diffuse emission reveals Mg-K (at 1.50 keV) and Si-K (at 2.04 keV)
lines of hydrogen-like species.
Such an ionization state also require a plasma temperature of $\sim 1$
keV.

The 0.9 keV component in M31 reminds us of the fact that two
large-scale diffuse Galactic X-ray emission phenomena, namely the
Galactic ridge emission (\cite{ridge_Koyama,ridge_Kaneda,ridge_Valinia})
and the Galactic bulge emission (\cite{bulge_Kokubun}), both involve a
soft component expressed by thin-thermal plasma emission of a temperature
0.6--0.8 keV.
These soft components in the Galactic ridge and bulge emission,
integrated over the whole Galaxy, exhibit a 0.5--10 keV luminosity of
$\sim 4 \times 10^{39}$ erg s$^{-1}$ and $\geq 7 \times 10^{37}$
erg s$^{-1}$, respectively.
Considering our limited data integration region around the M31 nucleus
and the measured luminosity, the 0.9 keV RS component of M31 may be
similar to the soft component of the Galactic diffuse X-ray  emission.
Although the origin of these Galactic soft components is still
unsettled, an assembly of old supernova remnants may be a possibility
(\cite{ridge_Koyama,ridge_Kaneda,ridge_Valinia}).
If so, the 0.9 keV RS component in M31 may be related to the past
supernova activity in M31.

\subsection{The Cooler RS Component}

The softest spectral component that appears in the energy range below
$\sim 0.8$ keV has been modeled by a thin-thermal emission with
a temperature $\sim 0.3$ keV, even though there may be a room for
alternative modelings.
Although the nucleus of M31 emits very soft X-rays, its 0.3--7 keV
luminosity, $\sim 4.0 \times 10^{37}$ erg s$^{-1}$
(\cite{M31_nucleus_Garcia}a), falls by an order of magnitude below that
of our softest component.
Furthermore, our softest component is clearly extended (subsection 3.5).
Therefore, contribution from the M31 nucleus is considered negligible.

We tentatively consider the softest component as coming from an
optically-thin plasma of a temperature $\sim 0.3$ keV.
This agrees with the Chandra and XMM-Newton results on M31, and on
NGC~4697 (\cite{NGC4697_Sarazin}).
One obvious candidate for this cooler RS component is warm inter-stellar
medium, as has been confirmed in some external galaxies (e.g.,
\cite{NGC4697_Sarazin}).

Another promising candidate is an assembly of stellar coronae.
A solar-type star has a typical coronal temperature of $\sim 0.3$ keV,
and an X-ray luminosity 5--7 orders of magnitude lower than its
bolometric luminosity (e.g., \cite{star_Pallavicini}).
The integrated bolometric luminosity of M31 is $\sim 1 \times 10^{44}$
erg s$^{-1}$ (\cite{galaxy_Tully}), and that from the central $12'$
region is estimated to be $\sim 4 \times 10^{43}$ erg s$^{-1}$.
The measured luminosity of the 0.3 keV RS component
(table~\ref{tab:luminosity}) thus becomes  $\sim 6 \times 10^{-6}$
when normalized to the stellar bolometric luminosity therein.
Since this value is close to those found for individual coronal sources,
the cooler RS component may be contributed significantly by the stellar
coronae, as well as by warm inter-stellar medium.

\appendix
\section*{Model of the LMXB spectra}

The X-ray spectra of luminous ($\gtrsim 2 \times 10^{37}$ erg s$^{-1}$)
LMXBs are described successfully with the two-component physical model
by \citet{LMXB_Mitsuda}, consisting of a softer and a harder components.
The softer one is a DBB model, which represents the integrated emission
from an optically-thick accretion disk around the nonmagnetized neutron
star (\cite{disk_Hoshi,disk_Inoue}).
The harder one is a $\sim 2$ keV BB model, which represents the
emission from the central neutron-star surface where the kinetic energy
of the accreting matter is thermalized.
The X-ray spectra of LMXBs are hence described by four parameters;
the innermost disk temperature $T_{\rm in}$ and the normalization of
the DBB component, and the temperature $T_{\rm BB}$ and the
normalization of the BB component.
Luminous Galactic LMXBs exhibit rather narrow scatter both in
$kT_{\rm in}$ and $kT_{\rm BB}$, typically $kT_{\rm in} =0.7-1.5$ keV
and $kT_{\rm BB} = 1.3-2.5$ keV.
The two constituent components usually cross over at energies of 3--7
keV (\cite{LMXB_Mitsuda}).
This model has been verified against the LMXB spectra taken with
various satellites, including Tenma (\cite{LMXB_Mitsuda}),
Ginga (\cite{LMXB_Makishima}a),
and ASCA (\cite{LMXB_Asai}).

\begin{figure}
\begin{center}
\FigureFile(80mm,80mm){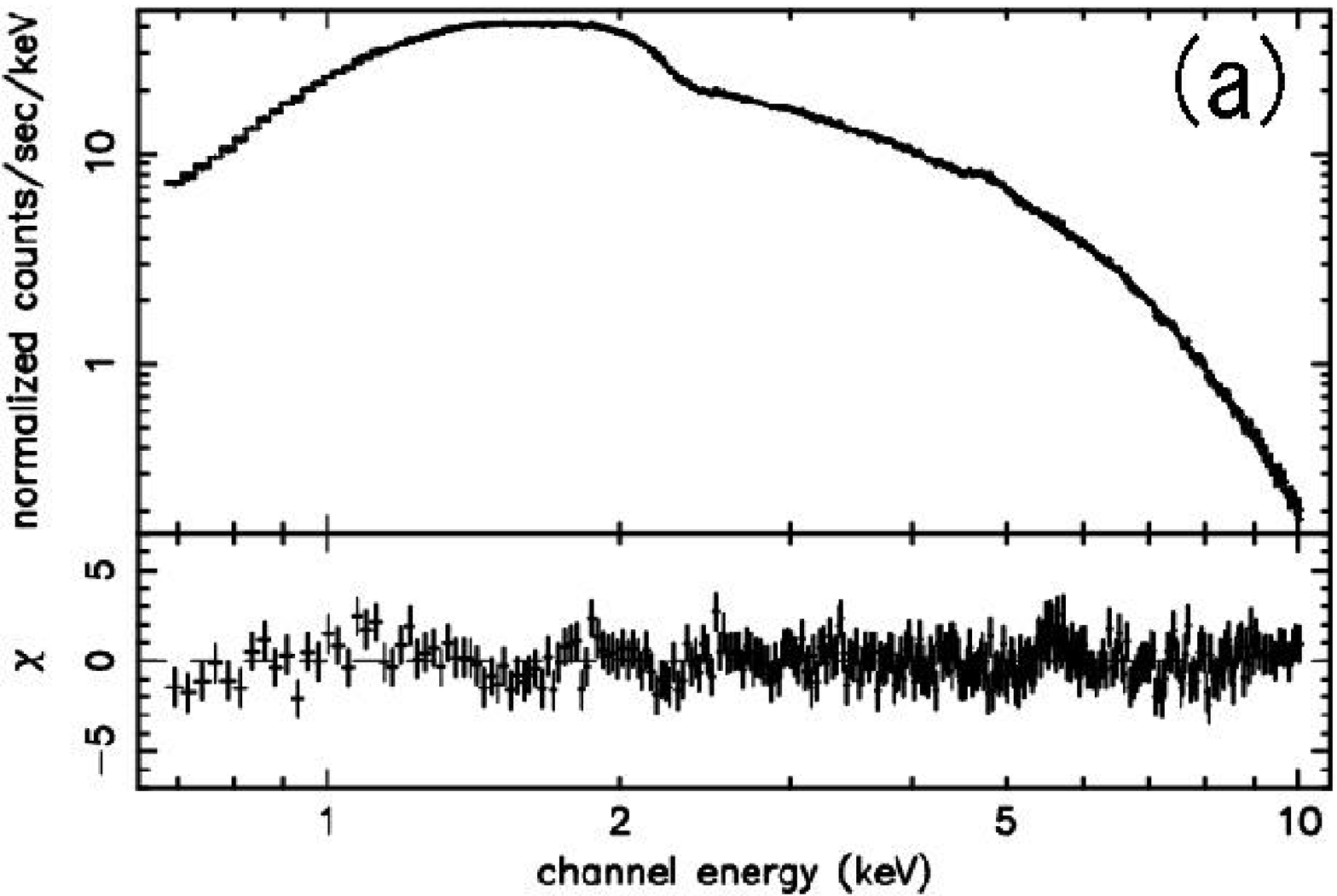}
\FigureFile(80mm,80mm){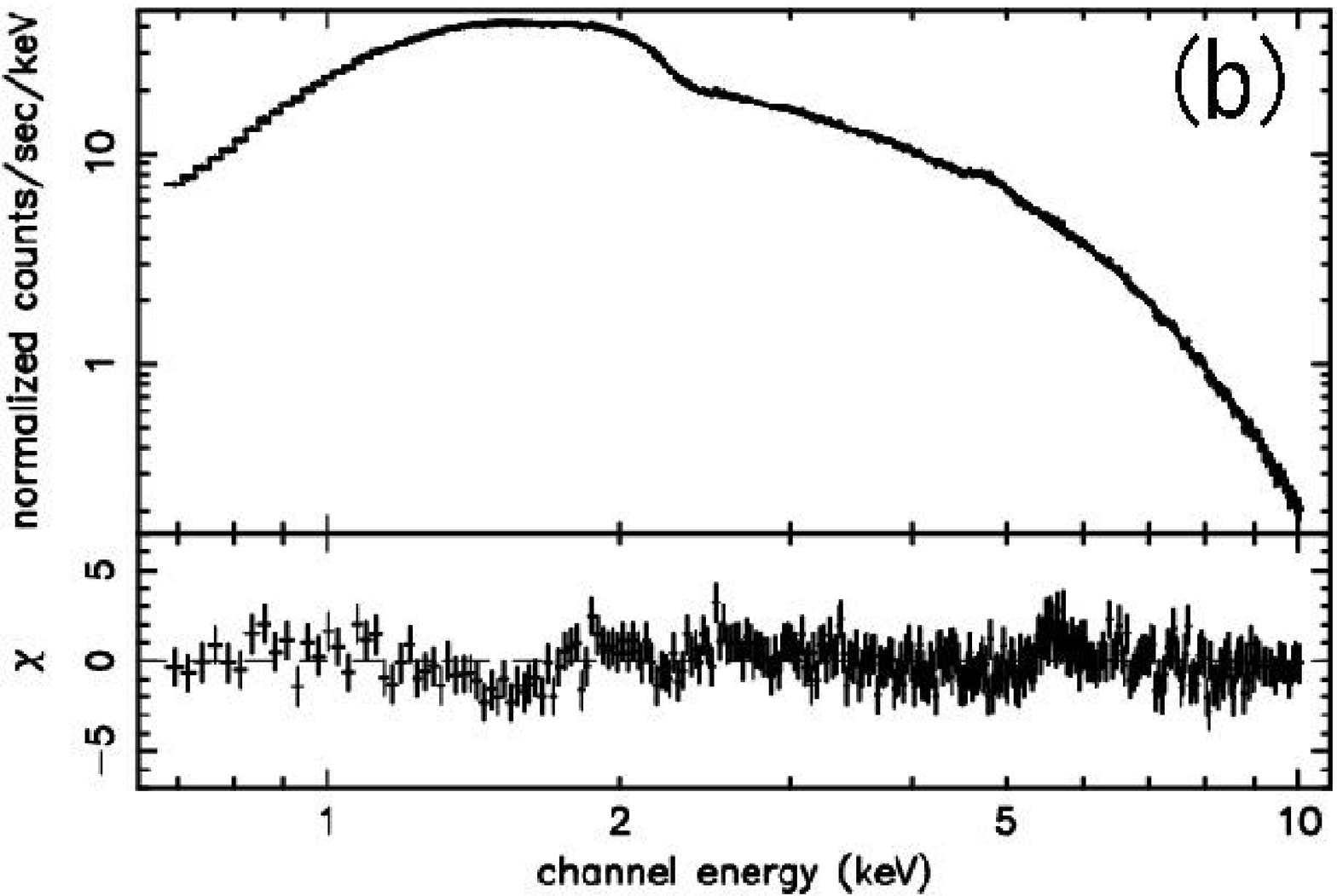}
\FigureFile(80mm,80mm){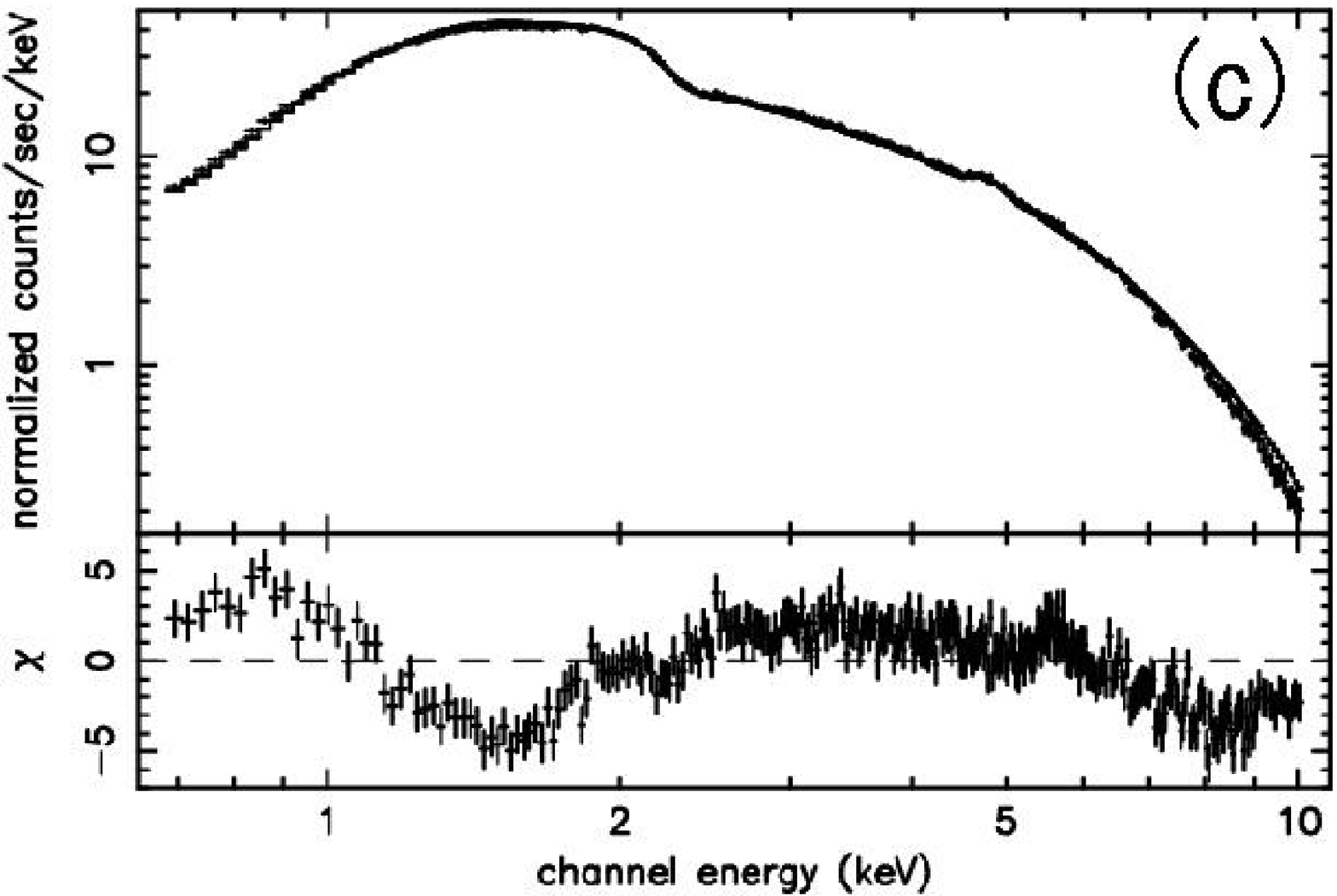}
\end{center}
\caption{Background-subtracted GIS spectrum of 4U 1820$-$30 in the
0.7--10 keV energy band.
The data are the same among the three panels.
The spectrum is fitted with (a) the LMXB model, (b) a single Bremss
model, and (c) a single PL model.}
\label{fig:4U1820}
\end{figure}

As a specific example, we fitted the spectrum of 4U $1820-30$ taken
with the ASCA GIS in October 1993.
This is a typical Galactic LMXB located in the globular cluster
NGC~6624, and exhibits a relatively low line-of-sight absorption.
It was hence used in \authorcite{M31_Makishima} as a comparison source.
Since the source is too bright for the SIS, we use the GIS data only.
We screened the GIS (GIS2 and GIS3) events as in section 2, to archive
a net exposure of $\sim 16$ ks.
The 0.7--10 keV count rate was $\sim 99$ c s$^{-1}$ per GIS detector,
corresponding to $\sim 120$ mCrab which is typical of 4U $1820-30$.
The spectrum has such a high statistics that the calibration
uncertainty becomes the dominant source of errors in the model fitting.
We therefore added a systematic error of 1\% to each bin of the GIS
spectrum.
For background, we used a source free region of the same observation
with the same radius.
Figure~\ref{fig:4U1820} shows the background subtracted GIS spectrum
without removing the instrumental response.

We fitted the 0.7--10 keV GIS spectrum of this LMXB with our LMXB
model, as well as the two conventional models, a single Bremss or a
single PL model.
We left the absorption column density free.
The LMXB model has reproduced the spectrum very well, in spite of the
very high signal statistics (figure~\ref{fig:4U1820}a,
table~\ref{tab:4U1820}).
On the contrary, the other two models showed poorer results
(figure~\ref{fig:4U1820}b and c, table~\ref{tab:4U1820}).
The Bremss fit, marginally acceptable, yielded
$N_{\rm H} \sim 2 \times 10^{21}$ cm$^{-2}$ and $kT_{\rm B} \sim 10$
keV.
These values are typical when an LMXB spectrum is approximated by a
Bremss model (\cite{LMXB_Makishima}a).
In this case, the absorption is thought to be artificial, required by
the slope of the Bremss continuum ($\sim 1.4$ in the energy range
$E \ll kT_{\rm B}$) which is steeper than that of the DBB model
($\sim \frac{2}{3}$ for $E \ll kT_{\rm in}$).
The temperature obtained through the Bremss fit to the M31 spectrum is
close to those for 4U 1820$-$30.

Although the LMXB fit is acceptable, the derived $N_{\rm H}$ is lower
than the Galactic value of $1.5 \times 10^{21}$ cm$^{-2}$ (from Einline
and W3nH).
This is probably due to a very weak ``soft excess'' component, emitted
by 4U~$1820-30$ itself or environment.
Actually, adding an RS model of a temperature $\sim 0.9$ keV has made
the LMXB fit even better, and this model is acceptable even when the
absorption is fixed at the Galactic value (table~\ref{tab:4U1820}).
The obtained flux in the 0.5--10 keV energy band of this soft
component ($2 \times 10^{-10}$ erg cm$^{-2}$ s$^{-1}$) is only
$\sim 2$\% of that of the LMXB component ($1 \times 10^{-8}$ erg
cm$^{-2}$ s$^{-1}$).
Therefore, this effect is much weaker than the excess soft X-ray
emission observed from M31.


\newpage

\newpage

\begin{table}
\caption{Spectral fit results for the central $12'$ region of
M31.$^{*,\dag}$}
\label{tab:LMXB}
\begin{center}
\begin{tabular}{llccccc}
\hline
\hline
 & & \multicolumn{2}{c}{LMXB} & 1st RS & 2nd RS \\
\cline{3-4}
\multicolumn{2}{l}{Range and Model} & $kT_{\rm in}^{\ddag}$
& $kT_{\rm BB}^{\ddag}$ & $kT_{\rm RS}^{\ddag}$ & $kT_{\rm RS}^{\ddag}$
& $\chi^2$/d.o.f.\ \\
\hline
\multicolumn{2}{l}{1.5--10 keV} \\
 & LMXB & $0.85\pm0.09$ & $1.9^{+0.3}_{-0.2}$ & $\cdots$
&  $\cdots$ & 208/221 \\
\hline
\multicolumn{2}{l}{0.8--10 keV} \\
 & LMXB & $0.47\pm0.03$ & $1.29^{+0.07}_{-0.06}$ & $\cdots$ &
$\cdots$ & 534/303 \\
 & LMXB+RS$^{\S}$ & $0.94^{+0.10}_{-0.09}$ & $2.0^{+0.4}_{-0.2}$ &
$0.79\pm0.03$ & $\cdots$ & 289/301 \\
\hline
\multicolumn{2}{l}{0.6--10 keV} \\
 & LMXB &$0.41^{+0.02}_{-0.03}$ & $1.20\pm0.05$ & $\cdots$
& $\cdots$ & 580/313 \\
 & LMXB+RS$^{\S}$ & $0.72\pm0.06$ & $1.7^{+0.1}_{-0.2}$ &
$0.33\pm0.03$ & $\cdots$ & 372/311 \\
 & LMXB+2RS$^{\S}$  & $0.9^{+0.2}_{-0.1}$ & $2.0^{+0.4}_{-0.3}$ &
$0.94^{+0.10}_{-0.07}$ & $0.28^{+0.03}_{-0.04}$ & 300/309 \\
\hline
\end{tabular}
\end{center}
\begin{footnotesize}
$^*$ $N_{\rm H}$ is fixed at the Galactic value of $6.7 \times 10^{20}$
cm$^{-2}$.\\
$^{\dag}$   All the errors are single-parameter 90\% confidence limits.
\\
$^{\ddag}$ Temperatures are all in the unit of keV.\\
$^{\S}$ The metal abundance of the RS component is fixed at 1.0 solar.
\end{footnotesize}
\end{table}

\begin{table}
\caption{Results of the LMXB+2RS model fits for the central $3'$ and
$6'$ regions in the 0.6--10 keV energy band.$^{*}$}
\label{tab:3'+6'}
\begin{center}
\begin{tabular}{cccccc}
\hline
\hline
 & \multicolumn{2}{c}{LMXB} & Hotter RS & Cooler RS & \\
\cline{2-3}
Radius & $kT_{\rm in}$ & $kT_{\rm BB}$ & $kT_{\rm RS}$ &  $kT_{\rm RS}$
& $\chi^2$/d.o.f.\ \\
\hline
$3'$ & $0.9^{+0.5}_{-0.2}$ & $1.6^{+2.1}_{-0.3}$ &
$0.95^{+0.10}_{-0.08}$ & $0.28\pm0.04$ & 105/130 \\
$6'$ & $0.9^{+0.2}_{-0.1}$ & $1.8^{+0.5}_{-0.2}$ &
$0.94^{+0.11}_{-0.08}$ & $0.29\pm0.04$  & 196/219 \\
\hline
\end{tabular}
\end{center}
\begin{footnotesize}
$^{*}$ The analysis conditions are the same as those for
table~\ref{tab:LMXB}.
\end{footnotesize}
\end{table}

\begin{table}
\caption{Luminosities of the three components for different
accumulation radii around the nucleus.$^{*,\dag}$}
\label{tab:luminosity}
\begin{center}
\begin{tabular}{clll}
\hline
\hline
Radius & LMXB component & Hotter RS component &  Cooler RS component \\
\hline
$3'$ & $~7.4^{+0.1}_{-0.2}$ ~~~~~~(1) & $0.8^{+0.2}_{-0.1}$ ~~~~~~(1)
& $0.7^{+0.2}_{-0.1}$ ~~~~(1) \\
$6'$ & $16.2\pm0.3$ ($2.20\pm0.06$) & $1.2^{+0.3}_{-0.2}$
($1.50\pm0.46$) & $1.5^{+0.1}_{-0.3}$ ($2.0\pm0.5$)\\
$12'$ & $26.0^{+0.4}_{-0.6}$ ~($3.5\pm0.1$) & $1.7^{+0.3}_{-0.4}$
~($2.0\pm0.6$) & $2.3^{+0.3}_{-0.2}$ ~($3.2\pm0.7$) \\
\hline
\end{tabular}
\end{center}
\begin{footnotesize}
$^*$ The luminosities are in the 0.5--10 keV band, in $10^{38}$ erg
s$^{-1}$.\\
$^{\dag}$ The numbers in parentheses are the ratios to the values for
the $3'$ radius.
\end{footnotesize}
\end{table}

\begin{table}
\caption{Alternative modelings of the 0.6--10 keV spectra of M31 for
the central $12'$ region.$^*$ }
\label{tab:bremss+power}
\begin{center}
\begin{tabular}{lccccc}
\hline
\hline
Models & $kT_{\rm B}^{\dag}/\Gamma^{\ddag}$ & $kT_{\rm RS}^{\dag}$ &
$kT_{\rm RS}^{\dag}$  & $\chi^2$/d.o.f.\ \\
\hline
Bremss$^{\S}$ & 5.9 & $\cdots$ & $\cdots$ & 683/315 \\
Bremss+RS & $7.4^{+0.5}_{-0.4}$ & $0.35\pm0.03$ & $\cdots$
& 364/313 \\
Bremss+2RS & $8.6^{+0.9}_{-0.5}$ & $0.90^{+0.12}_{-0.05}$ &
$0.28\pm0.04$ & 307/311 \\
\hline
PL & $1.79\pm0.02$ & $\cdots$ & $\cdots$ & $455/315$  \\
PL+RS & $1.71^{+0.03}_{-0.02}$ & $0.35^{+0.07}_{-0.04}$ & $\cdots$
& $362/313$  \\
PL+2RS & $1.68^{+0.02}_{-0.03}$ & $0.9\pm0.1$ &
$0.31^{+0.06}_{-0.07}$ & 348/311 \\
\hline
\end{tabular}
\end{center}
$^{*}$ The fitting conditions are the same as those for
table~\ref{tab:LMXB}.\\
$^{\dag}$ All temperatures are in keV.\\
$^{\ddag}$ $\Gamma$ is the PL photon index.\\
$^{\S}$ Errors are not shown because of the poor fit. \\
\end{table}

\begin{table}
\caption{Spectral fit results for the GIS spectrum of
4U $1820-30.^{*,\dag}$}
\label{tab:4U1820}
\begin{center}
\begin{tabular}{lccccccc}
\hline
\hline
Models & $N_{\rm H}^{\ddag}$ & $kT_{\rm in}^{\S}$ & $kT_{\rm BB}^{\S}$
& $kT_{\rm B}^{\S}$ & $\Gamma^{\P}$ & $kT_{\rm RS}^{\S}$ &
$\chi^2$/d.o.f.\ \\
\hline
LMXB & $6.5\pm0.7$ & $0.99\pm0.02$ &
$1.92^{+0.05}_{-0.03}$ & $\cdots$ & $\cdots$ &$\cdots$ & 283/297 \\
Bremss & $16.3\pm0.4$ & $\cdots$ & $\cdots$ &
$9.6^{+0.1}_{-0.2}$ & $\cdots$ & $\cdots$ & 329/299 \\
PL & 31 & $\cdots$ & $\cdots$ & $\cdots$ & 1.76 & $\cdots$ & 1463/299 \\
\hline
LMXB+RS$^{\|}$ & $15$ (fixed) & $0.92^{+0.02}_{-0.01}$ &
$1.86\pm0.03$ & $\cdots$ & $\cdots$ & $0.86\pm0.03$ &
307/296 \\
\hline
\end{tabular}
\end{center}
$^{\dag}$ A systematic error of 1\% is added to each data bin.\\
$^{\ast}$ Errors refer to 90\% confidence limits, but are not shown if
$\chi^2$/d.o.f.\ $> 2$.\\
$^{\ddag}$ $N_{\rm H}$ is in $10^{20}$ cm$^{-2}$.\\
$^{\S}$ All temperatures are in keV.\\
$^{\P}$ $\Gamma$ is the PL photon index.\\
$^{\|}$ The metal abundance is fixed at 1.0 solar.
\end{table}

\end{document}